\documentclass[twocolumn,aps,prd,superscriptaddress,nofootinbib]{revtex4-1}

\usepackage{graphicx}
\usepackage{color}
\usepackage{amsmath}
\usepackage{amssymb}
\usepackage{bm}
\usepackage{slashed}
\usepackage{epsfig}
\usepackage{amsfonts}
\usepackage{epstopdf}
\usepackage{bbm}
\usepackage{textcomp}
\usepackage{enumitem}
\usepackage{subfig}
\usepackage{hyperref}
\graphicspath{{fig/}}

\newcommand{\gev}{\mathrm{~GeV}}

\begin{document}
\title{Inclusive $\Upsilon(1S,2S,3S)$ photoproduction at the CEPC}
\author{Xi-Jie Zhan}
\email{zhanxj@ihep.ac.cn}
\affiliation{
	Institute of High Energy Physics (IHEP), Chinese Academy of Sciences (CAS),
	19B Yuquan Road, Shijingshan District, Beijing, 100049, China}
\affiliation{University of Chinese Academy of Sciences (UCAS), Chinese Academy of Sciences (CAS),
	19A Yuquanlu Road, Shijingshan District, Beijing, 100049, China}

\author{Jian-Xiong Wang}
\email{jxwang@ihep.ac.cn}
\affiliation{
	Institute of High Energy Physics (IHEP), Chinese Academy of Sciences (CAS),
	19B Yuquan Road, Shijingshan District, Beijing, 100049, China}
\affiliation{University of Chinese Academy of Sciences (UCAS), Chinese Academy of Sciences (CAS),
	19A Yuquanlu Road, Shijingshan District, Beijing, 100049, China}

\date{\today}

\begin{abstract}
The inclusive $\Upsilon(1S,2S,3S)$ photoproduction at the future Circular-Electron-Positron-Collider (CEPC) is studied based on the non-relativistic QCD (NRQCD) factorization formalism. Including the contributions from both direct and resolved photons, we present different distributions for $\Upsilon(1S,2S,3S)$ production and the results show there will be considerable events, which means that a well measurements on the $\Upsilon$ photoprodution could be performed to further study on the heavy quarkonium physics at electron-positron collider in addition to hadron colliders. This supplement study is very important to clarify the current situation of the heavy quarkonium production mechanism. 

\end{abstract}

\maketitle
\allowdisplaybreaks

\section{Introduction\label{sec:introduction}}

Study on heavy quarkonium plays a very important role to test quantum chromodynamics (QCD) on both perturbative and nonperturbative sides. Due to the heavy quark mass and hence the non-relativistic nature, non-relativistic QCD (NRQCD) factorization frame~\cite{Bodwin:1994jh} was proposed as a powerful tool to calculate its production and decay. The calculation is factorized into the products of the short distance coefficients (SDCs) and universal long distance matrix elements (LDMEs), and the SDCs are process-dependent and perturbatively double expansions in both the coupling constant $\alpha_s$ and the heavy quark relative velocity $v$, while the LDMEs can be fixed from experimental measurements. 

NRQCD predicts the process via color-octet mechanism and has achieved great successes, especially in $J/\psi$ production~\cite{Campbell:2007ws,Gong:2008ft,Butenschoen:2010rq,Ma:2010yw} and 
polarization~\cite{Butenschoen:2012px,Chao:2012iv,Gong:2012ug,Feng:2018ukp} at hadron colliders. As for bottomonium, due to the heavier mass of bottom quark, both the coupling constant $\alpha_s$  and the heavy quark relative velocity $v$ are smaller than charmonium case, which make it more suitable to be described by NRQCD. Some early investigations of bottomonium production can be found in Refs.\cite{Braaten:2000gw,Campbell:2007ws,Artoisenet:2008fc,Gong:2008hk,Gong:2010bk,Wang:2012is,Likhoded:2012hw} and references therein. The latest works on the full next-to-leading order(NLO) NRQCD studies of inclusive hadroproduction of $\Upsilon$ were done in Refs.\cite{Gong:2013qka,Han:2014kxa,Feng:2015wka,Feng:2020cvm}. In these papers, relative well agreements with experimental measurements were achieved but their fitted CO LDMEs show large difference when different schemes on NRQCD scale or different fitting strategies are applied. The situation indicates that further study and phenomenological testing of NRQCD is still an important task.
Besides hadron colliders, $e^+e^-$ colliders are also well place to study the physics of heavy quarkonium. There are advantages on both experimental and theoretical aspects~\cite{He:2019tig}. Experimentally, the backgrounds are less and cleaner for signal reconstruction, while theoretically the production mechanism is simpler and the uncertainties in calculation are smaller. At $e^+e^-$ collider, heavy quarkonium can be produced via two modes, i.e., $e^+e^-$ annihilation and $\gamma\gamma$ collision. 
The inclusive and exclusive charmonia production via $e^+e^-$ annihilation had been measured at B factories~\cite{Aubert:2001pd,Abe:2001za,Abe:2002rb,Abe:2004ww,Aubert:2005tj} and many theoretical works were done, seeing the review articles~\cite{He:2019tig,Lansberg:2019adr,Brambilla:2010cs}. Very recently, the calculations of charmonia production have marched on next-to-next-to-leading order(NNLO)~\cite{Feng:2015uha,Feng:2019zmt,Sang:2020fql}. As for the way of $\gamma\gamma$ collision, $J/\psi$ photoproduction had been measured at CERN LEP-II\cite{TodorovaNova:2001pt, Abdallah:2003du} and the leading order(LO) NRQCD calculation~\cite{Klasen:2001cu} in 2002 can describe the measurement. The NLO prediction~\cite{Butenschoen:2011yh} employing globally fitted LDMEs in 2011, however, was systematically overshot by the LEP-II data. It is worthy to note, however, that the uncertainties of LEP-II measurements on $J/\psi$ photoproduction are very large~\cite{He:2019tig}. 

As for the production of $\Upsilon$ mesons, there is so far no measurement yet at $e^+e^-$ colliders. 
The proposed Circular Electron Positron Collider (CEPC) \cite{CEPCStudyGroup:2018rmc,CEPCStudyGroup:2018ghi} can operate at different center of mass energy such as 91.2 GeV(Z pole), 161 GeV(WW threshold) and 240 GeV (Higgs factory). It's peak luminosity at 240$\gev$ is of order $10^{34}\mathrm{cm}^{-2}\mathrm{s}^{-1}$ and hence considerable heavy quarkonium events are expected. At the energy of 240$\gev$, the $\gamma\gamma$ collision mode (photoproduction) is dominant over the annihilation for heavy quarkonium production. The measurement on them can give precision results for different kinematics distributions and hopefully clarify the current predicament. Surely, it could also expand our knowledge of heavy quarkonium physics.

Therefore, to give an estimate and analysis on $\Upsilon$ production with roughly detector simulation at CEPC is very useful. In our previous work~\cite{Zhan:2020ugq}, we have investigated prompt $J/\psi$ photoproduction at the CEPC and presented promising results. There are also predictions of heavy quarkonium photoproduction at future $e^+e^-$ collider ILC ~\cite{Chen:2014xka,Sun:2015hhv}, where the photons are generated from laser backscattering(LBS) with electron and positron. The authors of Ref.~\cite{Chen:2014xka} calculated several heavy quarkonia photoproduction via color-singlet channels and their results show sizable $\Upsilon(1S)$ events yield.
In this work, based on the colliding photons from the electron positron bremsstrahlung, we investigate $\Upsilon(1S,2S,3S)$ photoproduction at the CEPC with considering both direct production and feed-down contributions from the heavier quarkonia.
In Section~\ref{sec:framework}, the basic theory framework for the calculation is outlined. The numerical results and analysis are shown in Section~\ref{sec:numerical}.  
Finally, a brief summary and conclusion are supplied in Section~\ref{sec:summary}.

\section{Photoproduction in NRQCD Framework\label{sec:framework}}

The colliding photons are from the electron positron bremsstrahlung, which is well described in Weiz\"acker-Williams approximation(WWA)~\cite{Frixione:1993yw},
\begin{eqnarray}
\label{eq:wwa}
f_{\gamma/e}(x) &=& \frac{\alpha}{2\pi}\Bigg[\frac{1 + (1 - x)^2}{x} {\rm log}\frac{Q^2_{max}}{Q^2_{min}} \nonumber\\
&&+2m_e^2x\left(\frac{1}{Q^2_{max}}
-\frac{1}{Q^2_{min}}\right)\Bigg],
\end{eqnarray}
where $\alpha=1/137$, is the electromagnetic fine structure constant, $Q^2_{min} = m_e^2 x^2/(1-x)$ and $Q^2_{max} = (E\theta_c)^2(1-x) + Q^2_{min}$ with $x = E_\gamma/E_e$. The maximum scattered angular cut $\theta_c$, is set as $32\mathrm{~mrad}$ to ensure the photon to be real, and $E=E_e=\sqrt{s}/2$ with $\sqrt{s}=240\gev$ at the CEPC.

In the NRQCD factorization, the SDCs stand for the production of an intermediate quark-antiquark pair which is in Fock state ($n={}^{2S+1}\!L^{[c]}_{J}$) with total spin $S$, orbital angular momentum $L$, total angular momentum $J$ and color-singlet (CS) $c=1$ or color-octet (CO) $c=8$. The LDMEs describe the probability of hadronization from the intermediate state to physical and colorless meson.
In the hard process, the photons from electron and positron can either collider directly or they can be resolved as hadronic components, which then collider with each other or photon. Under the factorization of NRQCD and the picture of WWA, the differential cross section of a hadron($H$) photoproduction is then formulated as the double convolution of the cross section of parton-parton (or photon) process and corresponding parton distribution functions,
\begin{eqnarray}
\label{tcs}
&&d\sigma(e^+e^-\to e^+e^-H+X)\nonumber\\
&&~=\int dx_1f_{\gamma/e}(x_1)\int dx_2f_{\gamma/e}(x_2)\nonumber\\
&&~~~~\times\sum_{i,j,k}\int dx_if_{i/\gamma}(x_i,\mu_f)\int dx_jf_{j/\gamma}(x_j,\mu_f)\nonumber\\
&&~~~~\times\sum_nd\sigma(ij\to b\overline{b}[n]+k)\langle{\cal O}^H[n]\rangle,
\end{eqnarray}
where $f_{i/\gamma}(x)$ is the Gl\"uck-Reya-Schienbein (GRS) parton distribution functions in photon~\cite{Gluck:1999ub},
$d\sigma(ij\to b\overline{b}[n]+k)$ are the differential partonic cross sections for $i,j=\gamma,g,q,\bar{q}$ and $k=g,q,\bar{q}$ with $q=u,d,s$. $b\overline{b}[n]$ is the intermediate $b\overline{b}$ Fock state with $n={}^3\!S_1^{(1)},{}^1\!S_0^{(8)},{}^3\!S_1^{(8)},{}^3\!P_J^{(8)}$ for $\Upsilon(mS)$ and $n={}^3\!P_J^{(1)},{}^3\!S_1^{(8)}$ for $H=\chi_{bJ}(mP)$ where $m=1,2,3$ and $J=0,1,2$. $\langle{\cal O}^H[n]\rangle$ is the LDME of $H$.

In addition to direct production, $\Upsilon$ mesons can also be produced via decays of heavier charmonia such as $\chi_{bJ}(mP)$. These feed-down contributions can be taken into account by multiplying their direct-production cross sections with the decay branching ratios to lighter ones, e.g.,
\begin{eqnarray}
\label{feeddown}
&&d\sigma^{\mathrm{total~}\Upsilon(1S)}=d\sigma^{\Upsilon(1S)}\nonumber\\
&&~~~~~~+\sum_{m,J}d\sigma^{\chi_{bJ}(mP)}Br(\chi_{bJ}(mP)\rightarrow \Upsilon(1S))\nonumber\\
&&~~~~~~+\sum_{m=2,3}d\sigma^{\Upsilon(mS)}Br(\Upsilon(mS)\rightarrow \Upsilon(1S)).
\end{eqnarray}

\section{Numerical Results\label{sec:numerical}}

The FDC package\cite{Wang:2004du} are used to generate the Fortran source for numerical calculation for all the related physics processes. In the calculations of sub parton-parton processes, the electromagnetic fine structure constant is set as $\alpha$=1/128 for the typical energy scale is of order $10\gev$ and one-loop running strong coupling constant $\alpha_s(\mu_r)$ is used.
The mass of bottom quark is approximately chosen as $m_b=m_H/2$ to conserve the gauge invariant of the hard-scattering amplitudes. The relevant quarkonia masses and branching ratios can be found in Refs.\cite{Tanabashi:2018oca,Zyla:2020}. As for Br($\chi_{bJ}(3P)\rightarrow \Upsilon(mS)$), we take the values in Table II of Ref.\cite{Han:2014kxa}.
The factorization scale($\mu_f$) and renormalization scale($\mu_r$) are set to be $\mu_f=\mu_r=\mu_0=\sqrt{4m_b^2+p_t^2}$ as the default choice and will vary independently from $\mu_0/2$ to $2\mu_0$ to estimate the uncertainties, and here $p_t$ is the transverse momentum of $H$ meson.
A shift $p_t^H\approx p_t^{H'}\times(M_H/M_{H'})$ is also used when considering the
kinematics effect from higher excited states.

\begin{table}[!htp]
	\begin{center}
		\footnotesize
		\begin{tabular*}{80mm}{c@{\extracolsep{\fill}}ccc}
			\hline\hline
			$\Upsilon(nS)$ & $|R_{\Upsilon(nS)}(0)|^{2}$ & $\chi_b(mP)$ & $|R'{\chi_{b(mP)}}(0)|^{2}$ \\[0.1cm]
			\hline
			1S & 6.477 GeV$^3$ & 1P & 1.417 GeV$^{5}$ \\
			2S & 3.234 GeV$^3$ & 2P & 1.653 GeV$^{5}$ \\
			3S & 2.474 GeV$^3$ & 3P & 1.794 GeV$^{5}$ \\
			\hline\hline
		\end{tabular*}
		\caption{\label{tab:csldmes}  Radial wave functions at the origin~\cite{Eichten:1995ch}.}
	\end{center}
\end{table}

\begin{table}[!htp]
	\begin{tabular}{{{c}cccc}}
		\hline\hline
		~state~&~Feng1~&~Feng2~&~Feng3~&~Han2016~
		\\\hline $\langle\mathcal{O}^{\Upsilon(1S)}({}^1\!S_0^{[8]})\rangle$  & $13.6$ & $10.1$ & $11.6$ & $13.7$
		\\\hline $\langle\mathcal{O}^{\Upsilon(1S)}({}^3\!S_1^{[8]})\rangle$  & $0.61$ & $0.73$ & 0.47 & $1.17$
		\\\hline $\langle\mathcal{O}^{\Upsilon(1S)}({}^3\!P_0^{[8]})\rangle/m_Q^2$  & $-0.93$ & $-0.23$  & $-0.49$ & $--$
		\\\hline $\langle\mathcal{O}^{\Upsilon(2S)}({}^1\!S_0^{[8]})\rangle$  & $0.62$ & $1.91$  & $-0.59$ & $6.07$
		\\\hline $\langle\mathcal{O}^{\Upsilon(2S)}({}^3\!S_1^{[8]})\rangle$  & $2.22$ & $1.88$  & $2.94$ & $1.08$
		\\\hline $\langle\mathcal{O}^{\Upsilon(2S)}({}^3\!P_0^{[8]})\rangle/m_Q^2$  & $-0.13$ & $-0.01$  & $0.28$ & $--$
		\\\hline $\langle\mathcal{O}^{\Upsilon(3S)}({}^1\!S_0^{[8]})\rangle$  & $1.45$ & $-0.15$  & $-0.18$ & $2.83$
		\\\hline $\langle\mathcal{O}^{\Upsilon(3S)}({}^3\!S_1^{[8]})\rangle$  & $1.32$ & $1.53$  & $1.52$ & $0.83$
		\\\hline $\langle\mathcal{O}^{\Upsilon(3S)}({}^3\!P_0^{[8]})\rangle/m_Q^2$  & $-0.27$ & $-0.02$  & $-0.01$ & $--$
		\\\hline $\langle\mathcal{O}^{\chi_{b0}(1P)}({}^3\!S_1^{[8]})\rangle$ & $0.94$  & $0.91$   & $1.16$ & $0.71$
		\\\hline $\langle\mathcal{O}^{\chi_{b0}(2P)}({}^3\!S_1^{[8]})\rangle$ & $1.09$  & $1.07$  & $1.50$ & $1.37$
		\\\hline $\langle\mathcal{O}^{\chi_{b0}(3P)}({}^3\!S_1^{[8]})\rangle$ & $0.69$  & $1.76$   & $1.92$ & $2.15$
		\\\hline\hline
	\end{tabular}
	\caption{\label{tab:coldmes} Different sets of CO LDMEs (in units of $10^{-2}$ GeV$^3$). The sets of Feng1(2,3) are taken from Table 2(3,4) of Ref.\cite{Feng:2015wka} and the set of Han2016 is taken from Ref.\cite{Han:2014kxa}.}
\end{table}

\begin{table}[!htp]
	\begin{tabular*}{80mm}{c@{\extracolsep{\fill}}ccc}
		\hline\hline
		~$\sqrt{S}$(GeV)~&~91.2~&~161~&~240~
		\\\hline 
		$~$  & CS,~NRQCD & CS,~NRQCD & CS,~NRQCD 
		\\\hline 
		$\sigma_{\Upsilon(1S)}$(fb)  & $0.88,10.99$ & $3.13,34.23$ & $6.34,67.66$ 
		\\\hline 
		$\sigma_{\Upsilon(2S)}$(fb)  & $0.32,3.75$ & $1.15,11.86$ & $2.43,23.75$ 
		\\\hline 
		$\sigma_{\Upsilon(3S)}$(fb)  & $0.20,1.56$ & $0.71,5.02$  & $1.51,10.13$
		\\\hline\hline
	\end{tabular*}
	\caption{\label{tab:tc} Total cross sections for $\Upsilon(1S,2S,3S)$ photoproduction at the CEPC with three typical collision energies. Here we have considered the feed-down contributions and take Han2016 LDMEs for NRQCD predictions.}
\end{table}

\begin{figure*} 
	\begin{center}
		\includegraphics[width=5.0cm]{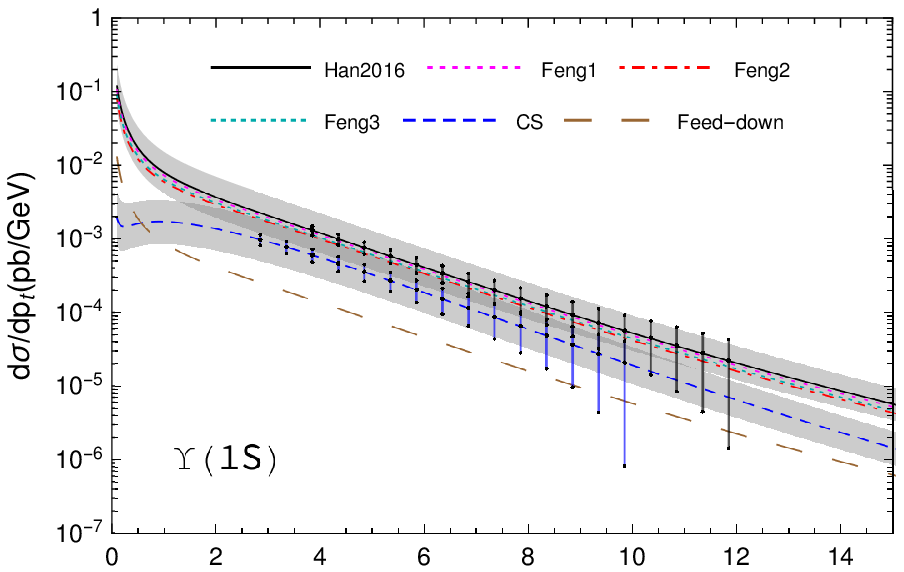} 
		\includegraphics[width=5.0cm]{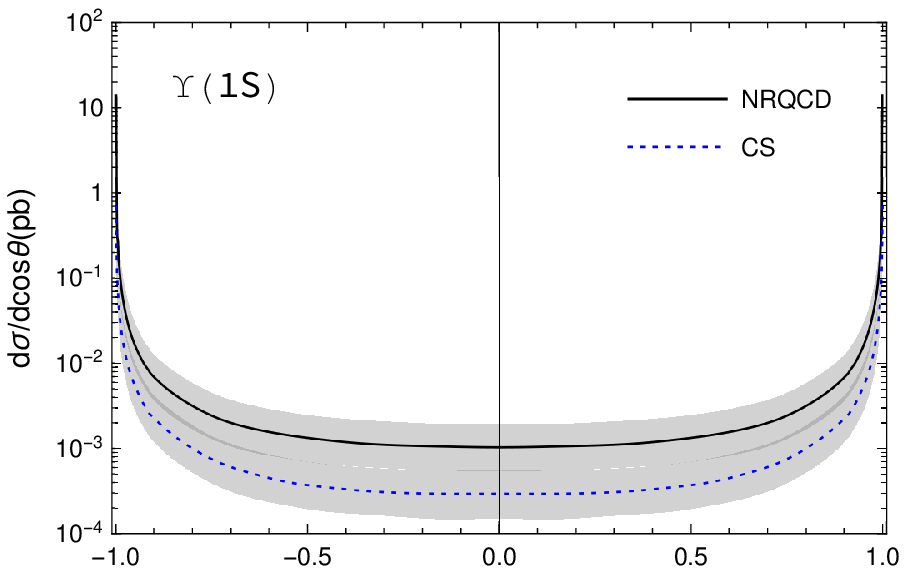} 
		\includegraphics[width=5.0cm]{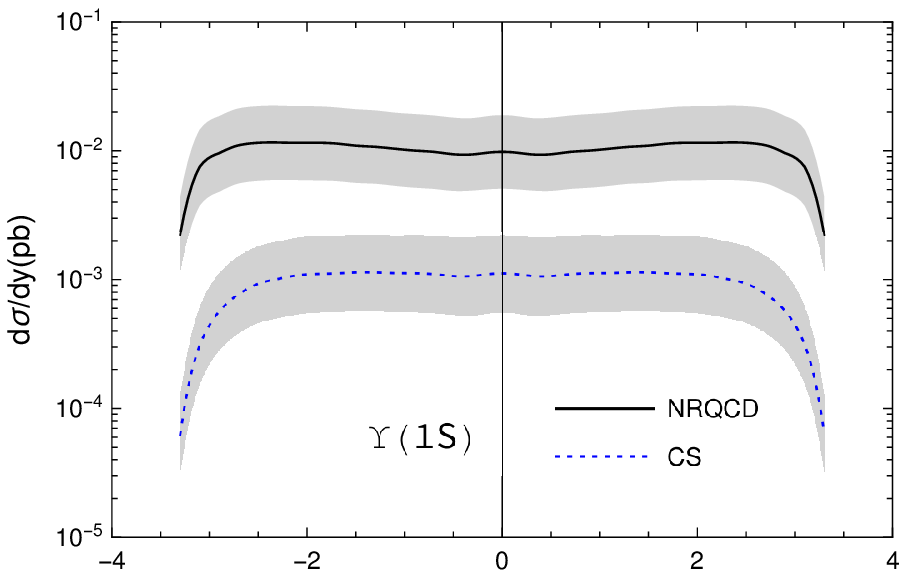} \\
		\includegraphics[width=5.0cm]{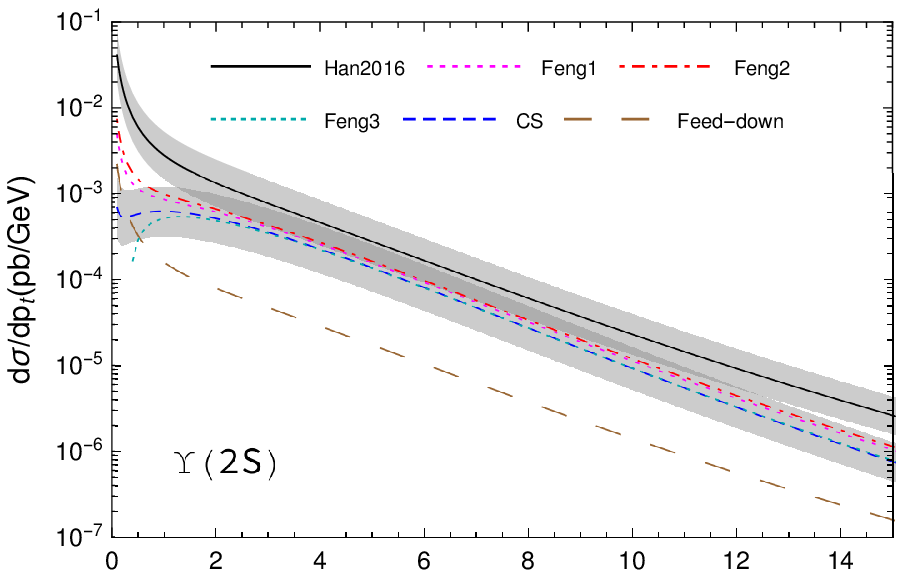}  
		\includegraphics[width=5.0cm]{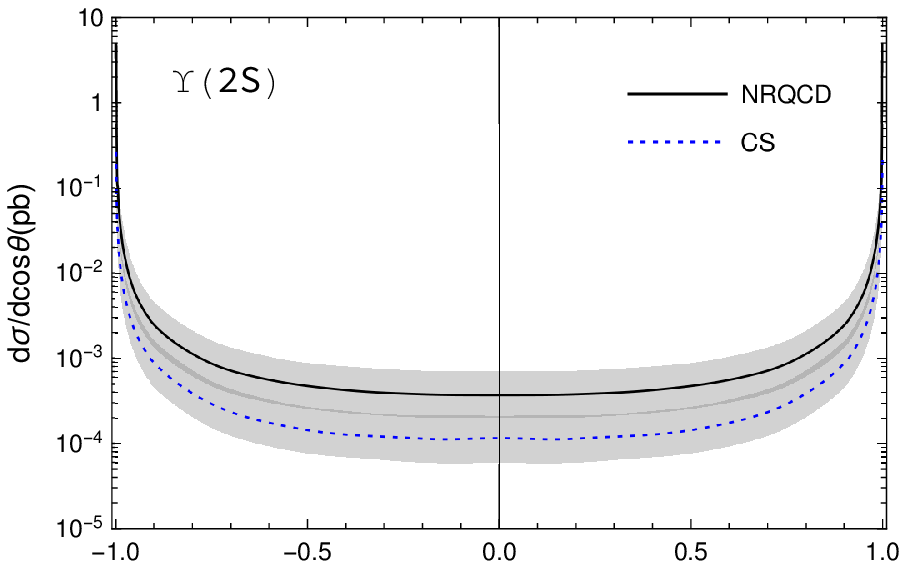}  
		\includegraphics[width=5.0cm]{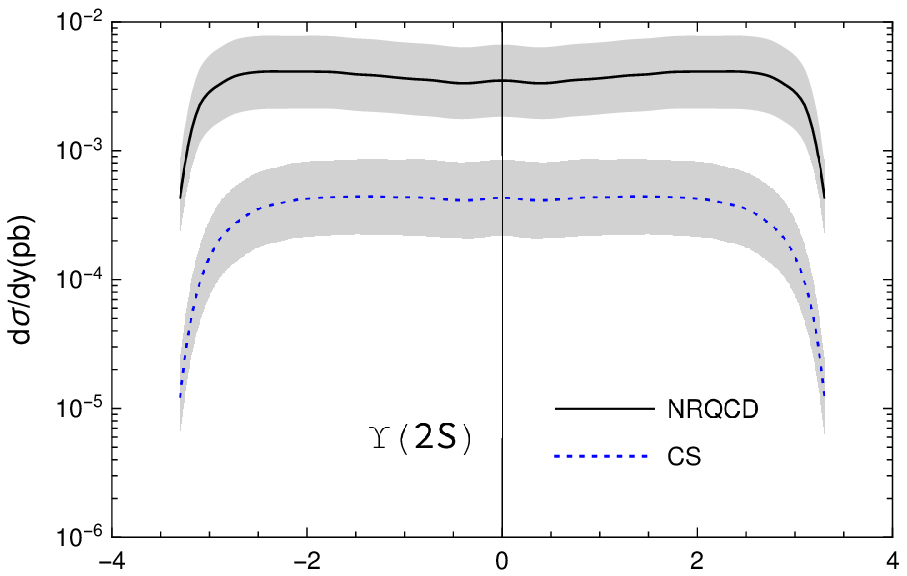} \\
		\includegraphics[width=5.0cm]{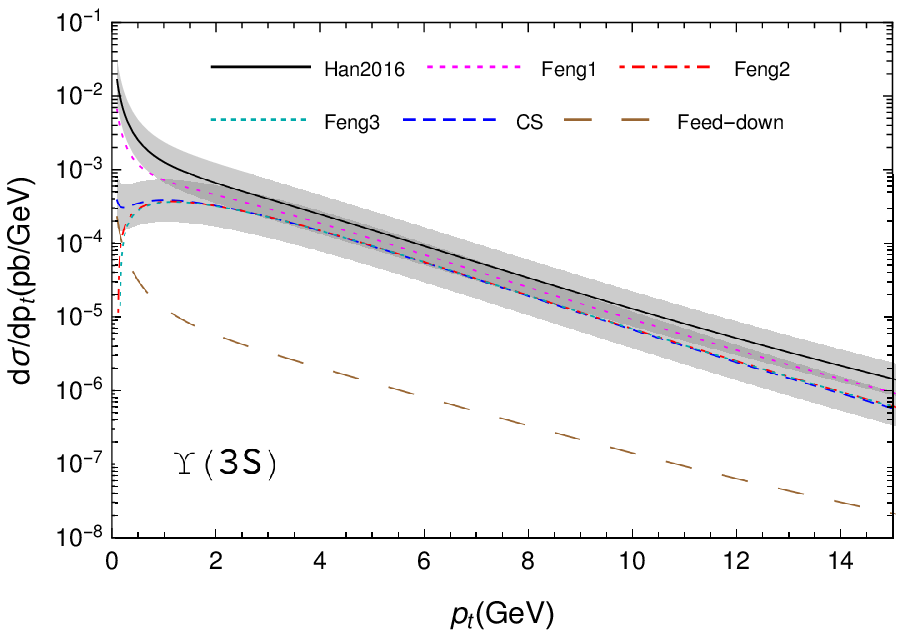}  
		\includegraphics[width=5.0cm]{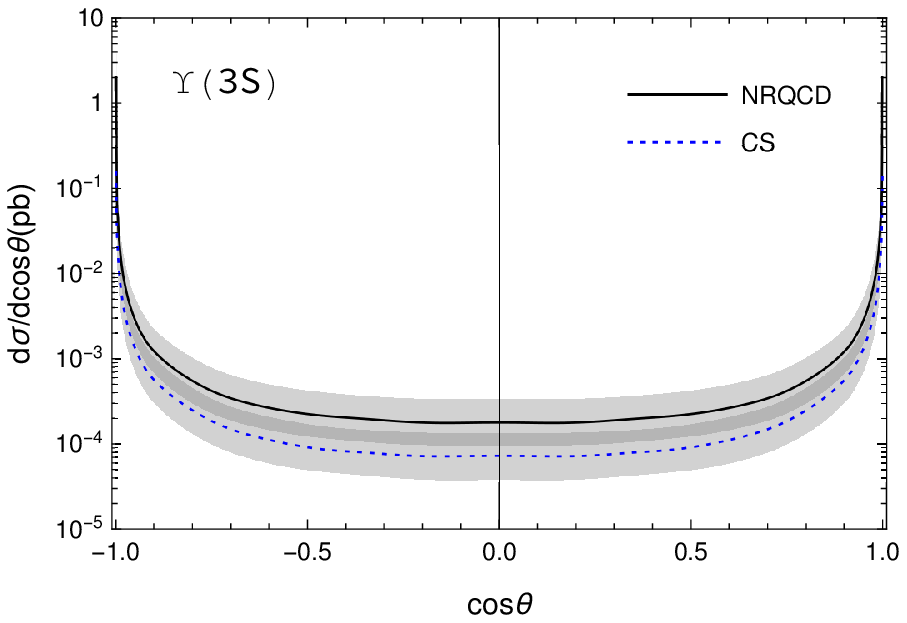}  
		\includegraphics[width=5.0cm]{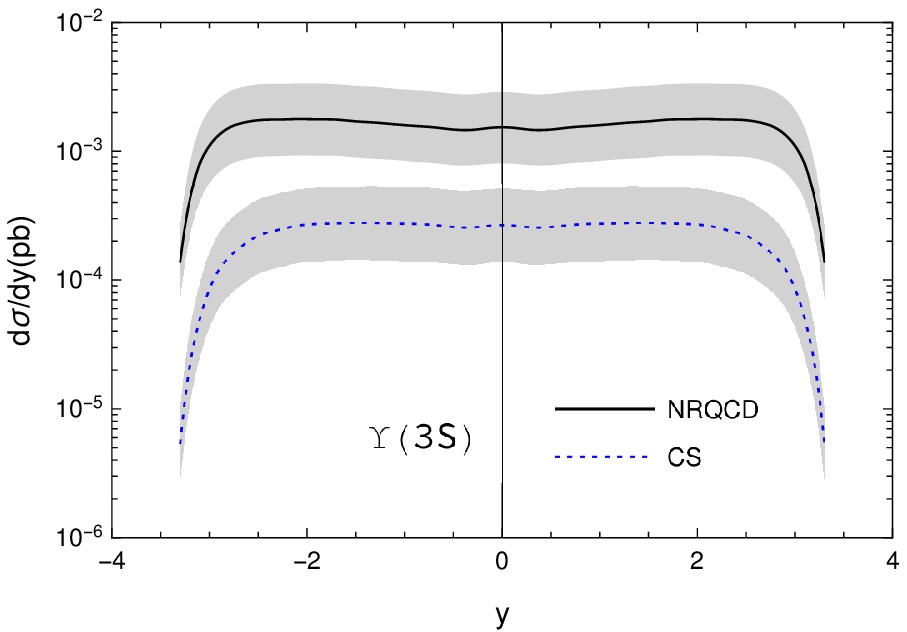} \\
	\end{center}
	\caption{\label{fig:distr1}
		The $p_t$(left), cos$\theta$(mid), and $y$(right) distribution of $\Upsilon$ photoproduction. The cos$\theta$ and $y$ plots only employ the CO LDMEs of Han2016. The light gray bands represent the theoretical uncertainties from $\mu_r$ and $\mu_f$ dependence, and the vertical lines in the $p_t$ distribution plot show the statistic error estimated from our simple detector simulation. Here only the uncertainties for color-singlet and CO LDMEs of Han2016 are shown.}
	\label{fig:lamda-def}
\end{figure*}

The color-singlet(CS) LDMEs are related to the wave functions at the origin by
\begin{eqnarray}
\langle{\cal O}^{\Upsilon(nS)}(^{3}S^{[1]}_{1})\rangle&=&\frac{9}{2\pi}|R_{\Upsilon(nS)}(0)|^{2}, \nonumber\\
\langle{\cal O}^{\chi_{bJ}(mP)}(^{3}P^{[1]}_{J})\rangle&=&\frac{3}{4\pi}(2J+1)|R'_{\chi_{b}(mP)}(0)|^{2}.
\end{eqnarray}
The wave functions at the origin can be obtained via potential model~\cite{Eichten:1995ch},
which are listed in Table~\ref{tab:csldmes}.

Several sets of color-octet(CO) LDMEs can be found in literatures and it is instructive to compare their predictions for the $\Upsilon$ photoproduction at $e^+e^-$ collision. We employ four different sets of CO LDMEs listed in Table~\ref{tab:coldmes}. The values of Feng1(2,3) are taken from Table 2(3,4) of Ref.\cite{Feng:2015wka} and the authors gave these three sets of CO LDMEs according to different fitting schemes. The set of Han2016 is taken from Ref.\cite{Han:2014kxa} where the authors decomposed the contribution of P-wave color-octet subprocesses into the linear combination of the two S-wave subprocesses and consequently they extract two linear combinations with three CO LDMEs which read,
\begin{eqnarray}
M^{\Upsilon}_{0,r_0}&=&\langle
O^{\Upsilon}(^1S_0^8)\rangle+\frac{r_0}{m^2_b}\langle
O^{\Upsilon}(^3P_0^8)\rangle,\label{eqn:M0}\\
M^{\Upsilon}_{1,r_1}&=&\langle
O^{\Upsilon}(^3S_1^8)\rangle+\frac{r_1}{m^2_b}\langle
O^{\Upsilon}(^3P_0^8)\rangle, \label{eqn:M1}
\end{eqnarray}
where $r_0$=3.8, $r_1$=-0.52, $M^{\Upsilon}_{0,r_0}=13.70\times10^{-2}$GeV$^3$ and $M^{\Upsilon}_{1,r_1}=1.17\times10^{-2}$GeV$^3$.

Table~\ref{tab:tc} lists the total cross sections for $\Upsilon(1S,2S,3S)$ photoproduction at three typical collision energies of the CEPC.
It shows that the cross sections increase with the growing of collision energy and the contributions of CO vastly dominate over CS.
The integrated luminosities per year of the CEPC are 8$~ab^{-1}$, 2.6$~ab^{-1}$ and 0.8$~ab^{-1}$ for collision energies 91.2~GeV, 161~GeV and 240~GeV respectively. The CEPC will run first for 7 years as a Higgs factory(240~GeV), followed by 2 years as a Super Z factory(91~GeV) and then 1 year of operation as a W factory(161~GeV). Therefore, enormous $\Upsilon$ mesons would be produced and by employing LDMEs of Han2016, for example, it predicts $7.86\times10^4$(91.2~GeV), $7.91\times10^4$(161~GeV) and $4.81\times10^4$(240~GeV) $\Upsilon(1S)$ mesons per year respectively. In the following discussion, we adopt $\sqrt{S}=240$~GeV for Higgs factory is the primary physics goal of the CEPC.

Fig.~\ref{fig:distr1} shows the $p_t$, cos$\theta$ and rapidity($y$) distributions of $\Upsilon$ photoproduction, where $\theta$ is the angle between $\Upsilon$ momentum and the $e^+e^-$ beam. Both cos$\theta$ and $y$ distributions are calculated under the cut $p_t\geq0.01\gev$. 
We vary $\mu_r$ and $\mu_f$ from $\mu_0/2$ to $2\mu_0$ to estimate the theoretical uncertainties. When taking $\mu_r=\mu_f$ and varying them simultaneously, it shows that their uncertainties cancel with each other to some extent. Hence we vary them independently and the largest uncertainties are obtained with an upper bound for $\mu_r=\mu_0/2,\mu_f=2\mu_0$ and a lower bound for $\mu_r=2\mu_0,\mu_f=\mu_0/2$, as shown by the light gray bands in Fig.~\ref{fig:distr1}. The major of the uncertainties are from the variation of $\mu_f$ in the GRS parton distribution functions of photon~\cite{Gluck:1999ub}.	
The $p_t$ distributions in Fig.~\ref{fig:distr1} show that different CO LDMEs in Table~\ref{tab:coldmes} don't give consistent predictions for $\Upsilon$ photoproduction, and Feng2 and Feng3 even give unphysical results for $\Upsilon(2S)$ and $\Upsilon(3S)$. This is the situation different from that of $\Upsilon$ hadroproduction\cite{Feng:2015wka}, where the results of these CO LDMEs sets show little difference although they have sizable differences themselves.
From the curves in Fig.~\ref{fig:distr1}, after considering the uncertainties, there is no significant difference between the NRQCD and color-singlet predictions for $p_t$ and $\cos\theta$ distributions. But they are distinguishable in $y$ plots. This indicates that the $y$ distribution maybe a better observable than $p_t$ and $\cos\theta$ to discriminate the color-octet and color-singlet mechanisms at the CEPC.

\begin{figure}
	\includegraphics[scale=0.7]{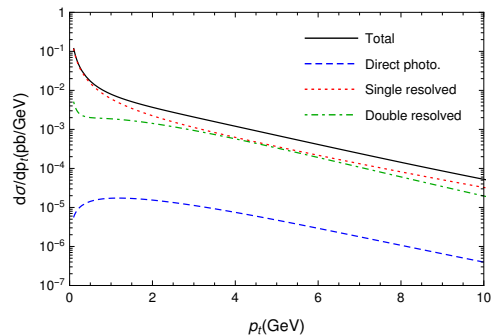}\\
	\caption{\label{fig:resolved}
		The $p_t$ distributions of the cross section from direct photoproduction and resolved photoproduction.}
\end{figure}

In Fig.~\ref{fig:distr1}, the feed-down contributions are shown by employing the CO LDMEs of Han2016(default choice). We can see that most of $\Upsilon$ mesons are produced directly. In the region $0.1\gev\leq p_t\leq10\gev$, for example, only (11, 5.6, 1.1)$\%$ of $\Upsilon(1S,2S,3S)$ are from decays of heavier charmonia respectively. The resolved channels are also dominated as shown in Fig.~\ref{fig:resolved}. As the reference, the $p_t$ distribution integrated from 0.1 to 10$\gev$, the direct, single-resolved and double-resolved channels account for 0.2$\%$, 80.4$\%$ and 19.4$\%$ of the NRQCD prediction, respectively.

\begin{figure}
	\includegraphics[scale=0.7]{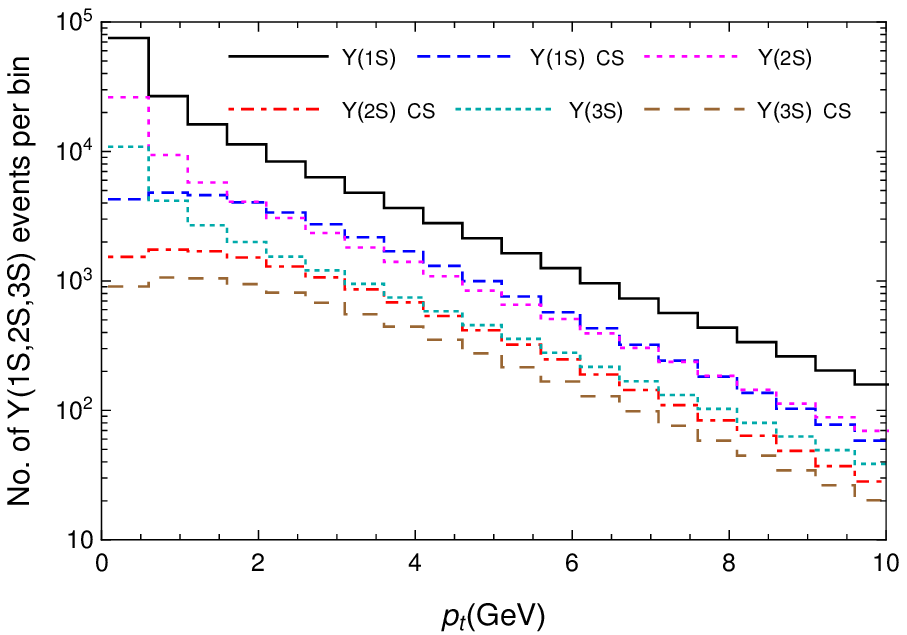}\\
	\includegraphics[scale=0.7]{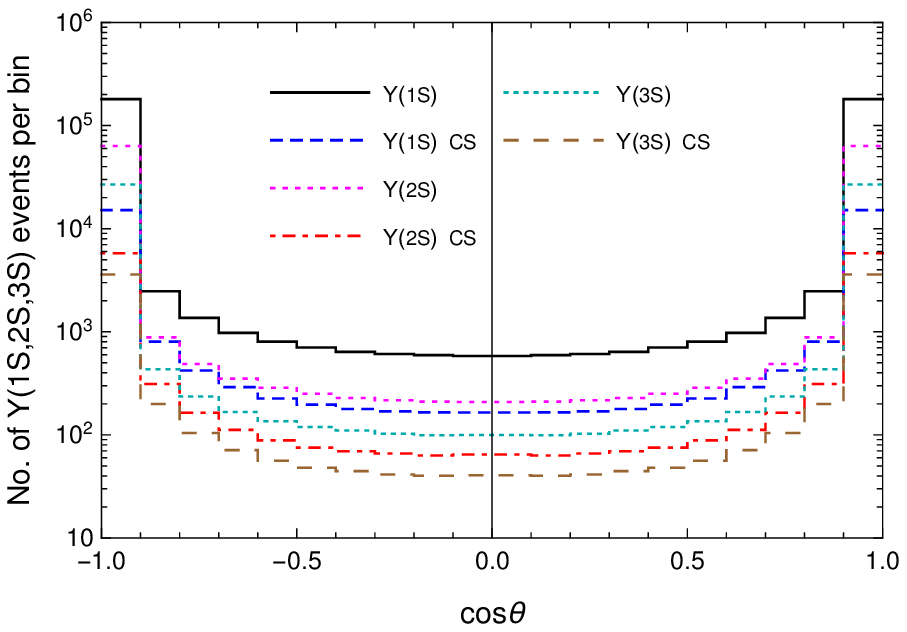}\\
	\includegraphics[scale=0.7]{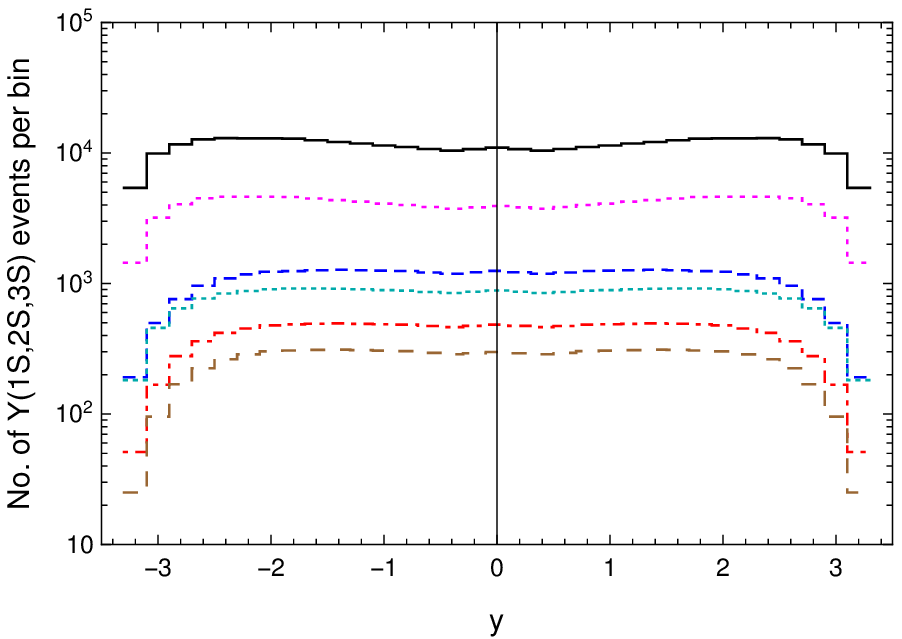}
	\caption{\label{fig:event}
		The event number distributions of $\Upsilon(1S,2S,3S)$. The bin widths are 0.5$\gev$ for $p_t$, 0.1 for $\cos\theta$ and 0.2 for y.
		The $y$ plots use same legends as $p_t$.}
\end{figure}

%

Fig.~\ref{fig:event} presents the number of $\Upsilon(1S,2S,3S)$ events distributions as function of $p_t$(upper), $\cos\theta$(middle) and $y$(lower) respectively for the integrated luminosity of the CEPC $5.6~\mathrm{ab}^{-1}$~\cite{CEPCStudyGroup:2018rmc}. The bin widths are 0.5$\gev$ for $p_t$, 0.1 for $\cos\theta$ and 0.2 for y. It shows that at the CEPC the number of events are considerable to discriminate between CS and NRQCD.

\begin{figure}[!h]
	\includegraphics[scale=0.7]{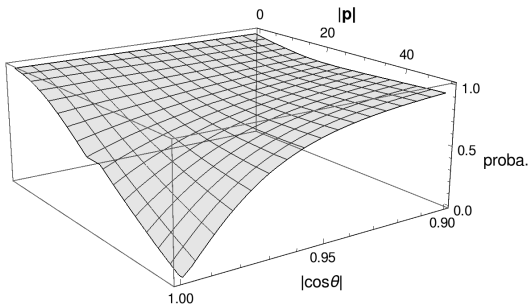}\\
	\caption{\label{fig:proba}
		The probability distribution of $\Upsilon(1S)$ with momentum $\textbf{p}$.}
\end{figure}

\begin{figure}[!h]
	\includegraphics[scale=0.7]{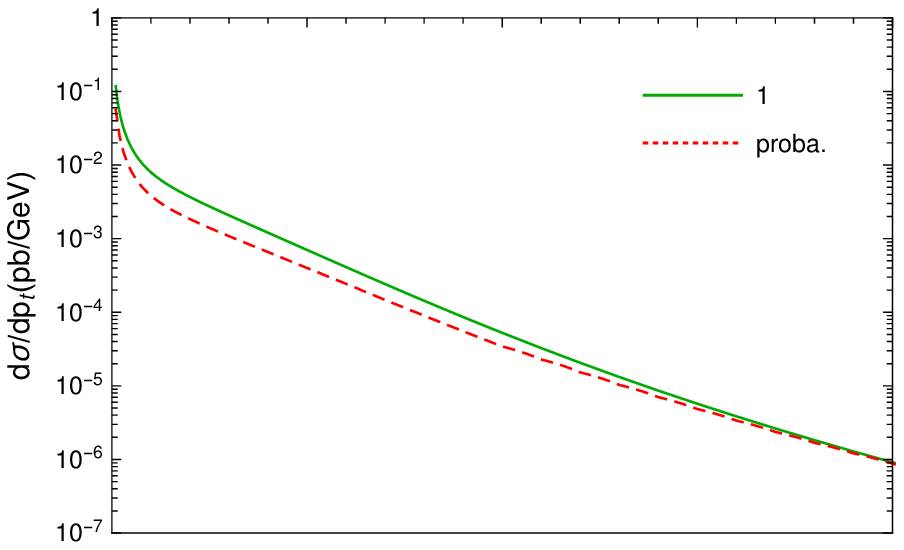}\\ \hspace{0.7mm}	
	\includegraphics[scale=0.7]{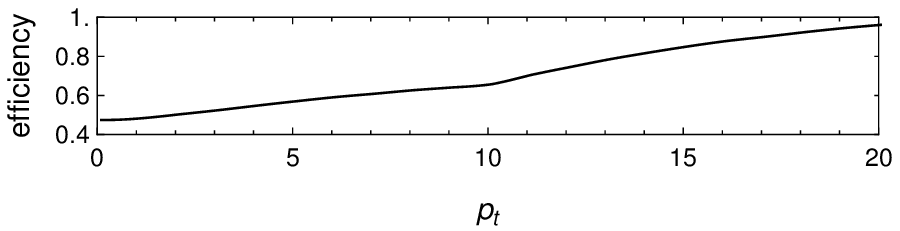}
	\includegraphics[scale=0.7]{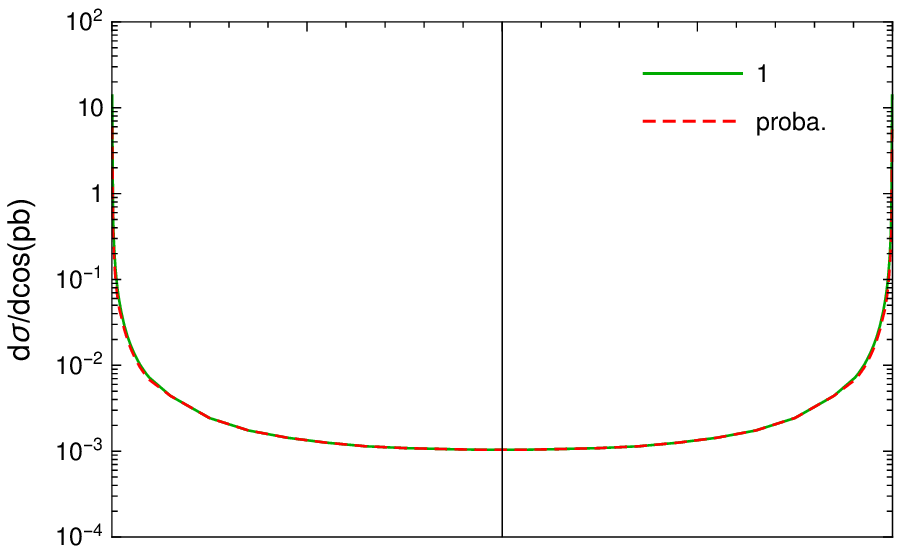}\\ \hspace{0.9mm}	
	\includegraphics[scale=0.7]{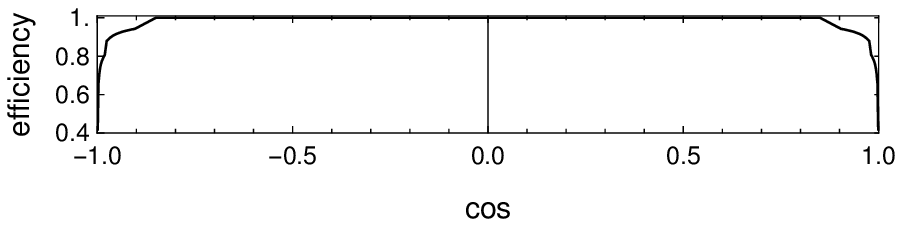}
	\includegraphics[scale=0.7]{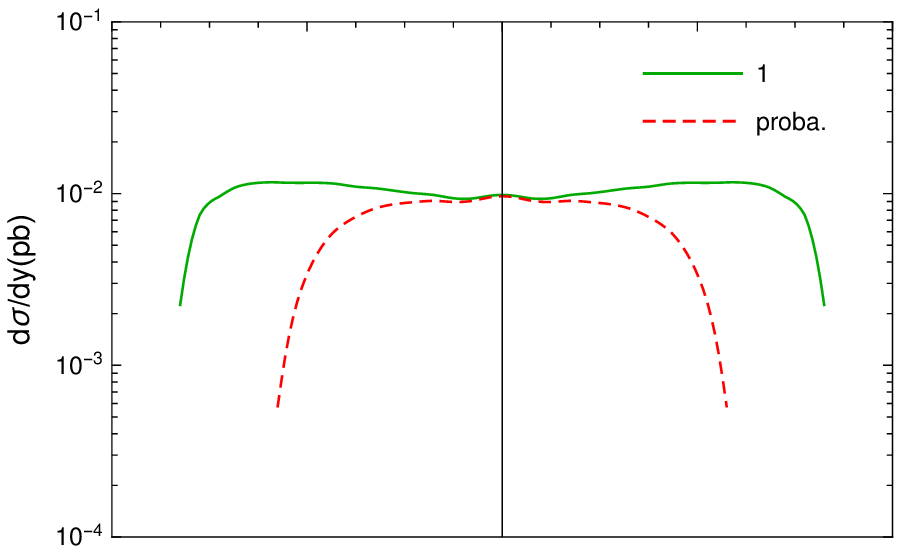}\\ \hspace{0.2mm}	
	\includegraphics[scale=0.7]{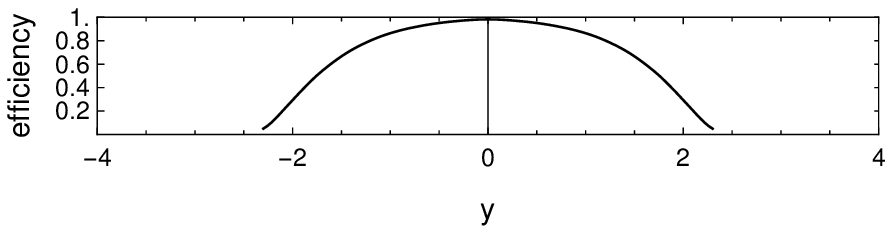}
	\caption{\label{fig:kinproba}
		The kinematic distributions of $\Upsilon(1S)$ photoproduction before(Line-1) and after(Line-proba.) considering the detection efficiency. The curves in the flat frames are the corresponding efficiencies.}
\end{figure}

According to the $\cos\theta$ plots in Fig.\ref{fig:event}, we see that most of $\Upsilon$ mesons are located in the closed beam region and actually more than 90$\%$ of $\Upsilon(1S)$ mesons are inside $|\cos\theta|\geq0.98$, which is the angular cut-off of experimental detection. In fact, however, $\Upsilon$ mesons decay almost immediately after their production at the colliding point. The $\mu^+\mu^-$ pair, for example, is used to reconstruct $\Upsilon$ meson in experiment and hence the probability of the $\mu^+\mu^-$ pair of being detected should be investigated. If both $\mu^+$ and $\mu^-$ are detected at the laboratory frame, then their parent $\Upsilon$ meson would be a valid event. So there is an issue of detection efficiency for $\Upsilon$. For simplicity we assume that, at the center-of-mass frame of $\Upsilon$, the $\mu^+\mu^-$ pair are isotropic in the whole $4\pi$ solid angle. Then we can easily calculate the probability of a $\Upsilon$ meson with given 4-momentum to be a valid event. Some brief derivations of this simple ``detector simulation" can be found in the Appendix of Ref~\cite{Zhan:2020ugq}. In Fig.~\ref{fig:proba} we plot the two-dimension distribution of the probability as function of magnitude of 3-momentum and $|\cos\theta|$ of $\Upsilon(1S)$. It shows that $\Upsilon(1S)$ meson, which has $|\cos\theta|\geq0.98$ but small $|\textbf{p}|$, still has the probability to be valid event. Fig.~\ref{fig:kinproba} shows the kinematic distributions both before(Line-1) and after(Line-proba.) considering the detection efficiency and here we only present the NRQCD results. The plots show that the efficiency gets larger as $p_t$ increases, which is reasonable as expected. The efficiency is close to one in most $\cos\theta$ region and $\Upsilon(1S)$ mesons with smaller $|\mathrm{y}|$ have larger probability of being valid. Consequently, there would be more valid events than those by directly using the experimental detecting angular cut to $\Upsilon$ mesons.

Considering the simple ``detector simulation" discussed above, the total detection efficiency for $\Upsilon(1S)$ are $83.68\%$(91.2~GeV), $74.05\%$(161~GeV) and $66.68\%$(240~GeV) respectively.
We further estimate the statistics uncertainties arising from the detection efficiency for the measurement on the $p_t$ distributions, which are shown in Fig.~\ref{fig:distr1} as the error bars at some $p_t$ points. We take the bin width $\Delta p_t=0.5\gev$ in the event counting. In small $p_t$ region, the uncertainties are smaller. Specifically, the uncertainties on the CS and NRQCD (employing Han2016 CO LDMEs) distributions are about 12.9$\%$ and 6.3$\%$ for $p_t=1\gev$ and about 26.5$\%$ and 18.9$\%$ for $p_t=5\gev$, respectively. In larger $p_t$ region, there would be less than one $\mu^+\mu^-$ pair in each bin for $p_t>10\gev$ and $p_t>12\gev$ for CS and NRQCD distributions respectively, and consequently larger bin width should be applied in experimental measurement.

\section{Summary\label{sec:summary}}

In this work, we have investigated the inclusive $\Upsilon(1S,2S,3S)$ photoproduction at the CEPC within the NRQCD framework at leading order, including the contributions from both direct and resolved photons. The dominate contribution is from the color-octet processes, and the decays of heavier bottomonia contribute 11$\%$ to the $\Upsilon(1S)$ production. Different kinematic distributions for both the production yield and the event number are presented based on the integrated luminosity $5.6~ab^{-1}$ of CEPC. It shows that the rapidity($y$) distribution is a better observable than that of $p_t$ and $\cos\theta$ to distinguish the color-singlet contribution and color-octet one. Under simple assumptions, the detecting efficiency of $\Upsilon$ is studied, and the results demonstrate that considerable $\Upsilon$ events could be reconstructed. Our results indicate that the measurement on $\Upsilon$ photoprodution at the CEPC can play an important role to find out whether or not only color-singlet mechanism contributes at $e^+e^-$ collider as the case of charmonia production at B factories, and to further test the color-octet mechanism in NRQCD and improve our understanding of the heavy quarkonium physics. We suggest that inclusive $\Upsilon$ photoproduction should be measured at the future CEPC. 

\begin{acknowledgments}
	
This work was supported by the National Natural Science Foundation of China with Grant No. 11975242 and the Key Research Program of Frontier Sciences, CAS, Grant No. Y7292610K1.

\end{acknowledgments}

\end{document}